\documentclass[a4paper]{jpconf}
\usepackage{graphicx}
\usepackage{hyperref}
\usepackage{amssymb,amsfonts} 
\usepackage{bm} 
\usepackage{mathrsfs} 
\usepackage{tikz}
\usetikzlibrary{calc,positioning,shapes,arrows,chains}

\hypersetup{pdftitle={SageManifolds},
  colorlinks=true
}

\newcommand{\soft}[1]{\textsf{#1}}
\newcommand{\code}[1]{\texttt{#1}}
\newcommand{\Sage}{\soft{Sage}}
\newcommand{\SM}{\soft{SageManifolds}}
\newcommand{\be}{\begin{equation}}
\newcommand{\ee}{\end{equation}}
\newcommand{\w}[1]{\bm{#1}}
\newcommand{\soutput}[1]{\textcolor{blue}{#1}\\[-0.8ex]\rule{\textwidth}{0.4pt}}

\begin{document}
\title{Tensor calculus with open-source software: \\
the SageManifolds project}

\author{Eric Gourgoulhon$^1$, Micha\l{} Bejger$^2$, Marco Mancini$^1$}

\address{$^1$ Laboratoire Univers et Th\'eories, UMR 8102 du 
CNRS, Observatoire de Paris, Universit\'e Paris Diderot,
92190 Meudon, France}

\address{$^2$ Centrum Astronomiczne im. M. Kopernika, ul. Bartycka 18,
00-716 Warsaw, Poland}

\ead{eric.gourgoulhon@obspm.fr, bejger@camk.edu.pl, marco.mancini@obspm.fr}

\begin{abstract}
The \SM{} project aims at extending the mathematics software system \Sage{} towards
differential geometry and tensor calculus. Like \Sage{}, \SM{} is free,
open-source and is based on the Python programming language.
We discuss here some details of the implementation, which relies 
on \Sage{}'s parent/element framework, and present a concrete example of use.
\end{abstract}

\tikzset{
base/.style={draw, thick, align=center},
native/.style={base, fill=cyan!30}, 
alg/.style={base, fill=red!40, rounded corners},
diff/.style = {base, fill=yellow!50, rounded corners},
dict/.style={base, fill=pink!40, draw=red},
tens/.style = {base, fill=yellow!25, align=left},
empty/.style={align=left},
legend/.style = {minimum width=2em, minimum height=1em},
native_legend/.style = {native, legend},
alg_legend/.style = {alg, legend},
diff_legend/.style = {diff, legend},
line/.style = {->, draw, thick, >=triangle 45}
}

\section{Introduction}

Computer algebra for general relativity (GR) has a long history, which started
almost as soon as computer algebra itself in the 1960s. 
The first GR program was \soft{GEOM}, written by J.G.~Fletcher
in 1965 \cite{Fletc67}. Its main capability was to compute the Riemann tensor
of a given metric. In 1969, R.A. d'Inverno developed \soft{ALAM}
(for \emph{Atlas Lisp Algebraic Manipulator}) and used it to compute
the Riemann and Ricci tensors of the Bondi metric.
According to \cite{Skea94}, 
the original calculations took Bondi and collaborators 6 months to finish,
while the computation with \soft{ALAM} took 4 minutes and yielded the 
discovery of 6 errors in the original paper. 
Since then, numerous packages have been developed: the reader is referred to \cite{MacCa02}
for a review of computer algebra
systems for GR prior to 2002, and to \cite{KorolKS13} for a more recent review 
focused on tensor calculus. 
It is also worth to point out the extensive list of
tensor calculus packages maintained by J. M. Martin-Garcia at \cite{xact_links}.


\section{Software for differential geometry}

Software packages for differential geometry and tensor calculus can be 
classified in two categories: 
\begin{enumerate}
\item Applications atop some general purpose computer algebra system. 
Notable examples are 
the \soft{xAct} suite \cite{Marti08} and \soft{Ricci} \cite{Ricci}, both
running atop \soft{Mathematica},
\soft{DifferentialGeometry} \cite{AnderT12} integrated into \soft{Maple},
\soft{GRTensorII} \cite{GRTensorII} atop \soft{Maple} and \soft{Atlas 2}
\cite{Atlas2} for \soft{Mathematica} and \soft{Maple}.
\item Standalone applications. Recent examples are \soft{Cadabra}  (field theory) \cite{Peete07},
\soft{SnapPy} (topology and geometry of 3-manifolds) \cite{SnapPy} and
\soft{Redberry} (tensors) \cite{BolotP13}.
\end{enumerate}
All applications listed in the second category are free software. In
the first category, \soft{xAct} and \soft{Ricci} are also free software, but
they require a proprietary product, the source code of which is closed (\soft{Mathematica}).

As far as tensor calculus is concerned, the above packages can be distinguished by 
the type of computation that they perform:
abstract calculus (\soft{xAct/xTensor}, \soft{Ricci}, \soft{Cadabra}, \soft{Redberry}),
or component calculus (\soft{xAct/xCoba}, \soft{DifferentialGeometry}, \soft{GRTensorII},
\soft{Atlas 2}). 
In the first category, tensor operations such as contraction or covariant differentiation 
are performed by manipulating the indices themselves rather than the components 
to which they correspond. In the second category, vector frames are explicitly 
introduced on the manifold and tensor operations are carried out on the components 
in a given frame.


\section{An overview of Sage}

\Sage{} \cite{sage} is a free, open-source mathematics software system, which is
based on the Python programming language. It makes use of over 90 open-source packages, 
among which are \soft{Maxima} and \soft{Pynac} (symbolic calculations),
\soft{GAP} (group theory), 
\soft{PARI/GP} (number theory), \soft{Singular} (polynomial computations), 
and \soft{matplotlib} (high quality 2D figures). 
\Sage{} provides a uniform Python interface to all these packages; however, 
\Sage{} is much more than a mere interface: it contains a large and increasing part of 
original code (more than 750,000 lines of Python and Cython, involving 5344 classes). 
\Sage{} was created in 2005 by W. Stein \cite{SteinJ05} and since
then its development has been sustained by more than a hundred researchers
(mostly mathematicians). Very good introductory textbooks about \Sage{} are
\cite{JoyneS14,Zimme13,Bard15}. 
 
Apart from the syntax, which is based on a popular programming language and not
a custom script 
language, a difference between \Sage{} and, e.g., \soft{Maple} or \soft{Mathematica}
is the usage of the \emph{parent/element pattern}. This framework more closely
reflects actual mathematics. 
For instance, in \soft{Mathematica}, all objects 
are trees of symbols and the program is essentially a set of 
sophisticated rules to manipulate symbols. On the contrary, in \Sage{}
each object has a given type (i.e. is an instance of a given
Python class\footnote{Let us
recall that within an object-oriented programming language (as Python),
a \emph{class} is a structure to declare and store the
properties common to a set of objects. These properties 
are data (called 
\emph{attributes} or \emph{state variables}) and functions acting 
on the data (called \emph{methods}). A specific realization of an object 
within a given class is called an \emph{instance} of that class.}), 
and one distinguishes \emph{parent} types, which model mathematical
sets with some structure (e.g. algebraic structure), from \emph{element} types,
which model set elements. Moreover, each parent belongs to some 
dynamically generated class that encodes informations 
about its \emph{category}, in the mathematical sense of the word
(see \cite{sage_categories} for a discussion of \Sage{}'s category framework).
Automatic conversion rules, called \emph{coercions},
prior to a binary operation, e.g. $x+y$ with $x$ and $y$ having different 
parents, are implemented.


\section{The SageManifolds project}

\subsection{Aim and scope}

\Sage{} is well developed in many areas of mathematics but 
very little exists for differential geometry and tensor calculus.
One may mention differential forms defined on a fixed coordinate patch,
implemented by J.~Vankerschaver \cite{sage_diff_form},
and the 2-dimensional parametrized surfaces of
the 3-dimensional Euclidean space recently added
 by M.~Malakhaltsev and J.~Vankerschaver
\cite{sage_param_surf}. 

The aim of \SM{} \cite{SM} is to introduce smooth manifolds and tensor fields
in \Sage{}, with the following requirements:
(i) one should be able to introduce various coordinate charts
on a manifold, with the relevant transition maps; (ii)
tensor fields must be manipulated as such and not through 
their components with respect to a specific (possibly coordinate) vector frame. 

\begin{figure}
\begin{center}
\begin{tikzpicture}[font=\small, node distance=0.5cm, minimum
height=2em, auto]

\node[native]
(unique_represenation){UniqueRepresentation};

\node[native, right=of unique_represenation]
(parent){Parent};

\coordinate (Middle) at ($(unique_represenation)!0.5!(parent)$);

\node[diff, below=1.5cm of Middle]
(domain) {ManifoldSubset\\ {\scriptsize {\it element:} ManifoldPoint}};

\path[line] (domain) -- (unique_represenation);
\path[line] (domain) -- node [near start, yshift=-1em, xshift=6em]
{\footnotesize {\it category:} Sets} (parent);

\node[diff, below=of domain]
(opendomain) {ManifoldOpenSubset};

\path[line] (opendomain) -- (domain);

\node[diff, below=of opendomain]
(manifold) {Manifold};

\path[line] (manifold) -- (opendomain);

\node[diff, below=of manifold]
(submanifold) {Submanifold};
\path[line] (submanifold) -- (manifold);

\node[diff, right=of submanifold]
(realline) {RealLine};
\path[line] (realline) -- (manifold);
\node[native, right=1.5cm of parent]
(element){Element};

\node[diff, below=1.225cm of element]
(point){ManifoldPoint};

\path[line] (point) -- (element);


\node[native_legend, left=5cm of opendomain]
(native_legend){};
\node[empty, right=0.5em of native_legend]
{Native \soft{Sage} class};

\node[diff_legend, below=1.em of native_legend]
(diff_legend){};
\node[empty, right=0.5em of diff_legend]
{\soft{SageManifolds} class\\ \footnotesize (differential part)};

\end{tikzpicture}
\end{center}
\caption{\label{f:domain_classes} Python classes for 
smooth manifolds (\code{Manifold}), generic subsets of them 
(\code{ManifoldSubset}), open subsets of them (\code{ManifoldOpenSubset})
and points on them (\code{ManifoldPoint}).}
\end{figure}
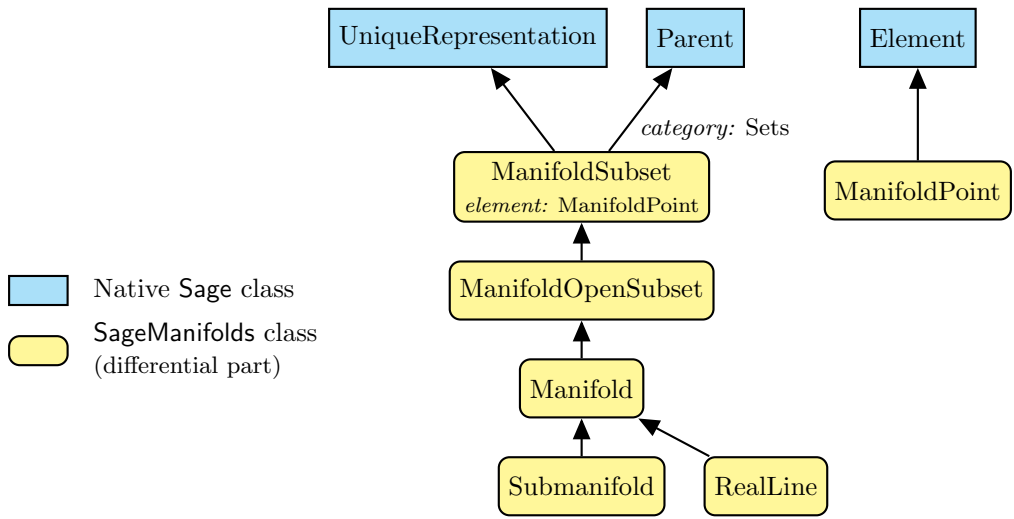

Concretely, the project amounts to creating new Python classes,
 such as 
\code{Manifold}, \code{Chart}, \code{TensorField} or \code{Metric},
to implement them within the parent/element pattern and to 
code mathematical operations as class methods.
For instance the class \code{Manifold}, devoted to real smooth manifolds,
is a parent class, i.e. it inherits from \Sage{}'s class \code{Parent}.
On the other hand, the class devoted to manifold points, \code{ManifoldPoint}, 
is an element class and therefore inherits from \Sage{}'s class 
\code{Element}.
This is illustrated by the inheritance diagram of Fig.~\ref{f:domain_classes}.
In this diagram, each class at the base of some arrow is a subclass (also
called \emph{derived class}) of the class at the arrowhead.
Note however that the actual type of a parent is a dynamically generated
class taking into account the mathematical category to which it belongs. 
For instance, the actual type of a smooth manifold is not \code{Manifold}, but a
subclass of it named \code{Manifold\_with\_category}, 
reflecting the fact that \code{Manifold} is
declared in the category of \code{Sets}\footnote{A tighter category would be
topological spaces, but such a category has been not implemented in \Sage{} yet.}.
Note also that the class \code{Manifold} inherits from \Sage{}'s class
\texttt{UniqueRepresentation}, which ensures that there is a unique
manifold instance for a given dimension and given name. 

\subsection{Implementation of charts}

Given a smooth manifold $\mathcal{M}$ of dimension $n$, a coordinate chart 
on some open subset $U\subset\mathcal{M}$ is implemented in \SM{} 
via the class \code{Chart}, whose main data is 
a $n$-tuple of \Sage{} symbolic variables $(x_1,\ldots,x_n)$, each of 
them representing a coordinate.
In general, more than one (regular) chart is required to cover the entire manifold.
For instance, at least 2 charts are necessary for the $n$-dimensional sphere 
$\mathbb{S}^n$ ($n\geq 1$) and the torus $\mathbb{T}^2$ and 3 charts for the real projective plane
$\mathbb{RP}^2$ (see Fig.~\ref{f:boy} below). 
Accordingly, \SM{} allows for an arbitrary number of charts. 
To fully specify the manifold, one shall also provide the transition maps
(changes of coordinates) on
overlapping chart domains (\SM{} class \code{CoordChange}).

\subsection{Implementation of scalar fields}

A \emph{scalar field} on manifold $\mathcal{M}$ is a smooth mapping
\be
    \begin{array}{lcll}
    f: & U\subset \mathcal{M}&\longrightarrow &\mathbb{R} \\
       & p & \longmapsto  & f(p) ,
    \end{array}
\ee
where $U$ is some open subset of $\mathcal{M}$.
A scalar field has different coordinate representations $F$, $\hat F$, etc. 
in different charts $X$, $\hat X$, etc. defined on $U$:
\be
    f(p) = 
F(\underbrace{x^1,\ldots, x^n}_{\mbox{coord. of $p$}\atop\mbox{in chart $X$}}) 
= {\hat F}(\underbrace{{\hat x}^1,\ldots, {\hat x}^n}_{\mbox{coord. of $p$}\atop\mbox{in chart $\hat X$}})
= \ldots
\ee
These representations are 
stored in some attribute of the class \code{ScalarField}, namely a 
Python dictionary\footnote{A \emph{dictionary}, also known as \emph{associative array}, is a 
data structure that generalizes the concept of array in the sense that the
key to access to some element is not restricted to an integer or a tuple of integers.} 
whose keys are the various charts defined on $U$:
\be \label{e:f_express}
 f.\mbox{\texttt{\_express}} = \left\{ X: F,\ \hat X: \hat F, \ldots \right\} .
\ee
Each representation $F$ is an instance of the class \code{FunctionChart}, 
which resembles \Sage{} native symbolic functions, but involves 
automatic simplifications in all arithmetic operations. 

\begin{figure}
\begin{center}
\begin{tikzpicture}[font=\small, node distance=0.5cm, minimum height=2em, auto]

\node[native]
(unique_represenation){UniqueRepresentation};

\node[native, right=of unique_represenation]
(parent){Parent};

\coordinate (Middle) at ($(unique_represenation)!0.5!(parent)$);

\node[diff, below=1.5cm of Middle]
(scalar_field_algebra)
{ScalarFieldAlgebra\\
{\scriptsize {\it ring:} SR}\\
{\scriptsize {\it element:} ScalarField}};

\path[line] (scalar_field_algebra) -- (unique_represenation);
\path[line] (scalar_field_algebra) -- node [near start, yshift=-1em,
xshift=13em]
{\footnotesize {\it category:} CommutativeAlgebras} (parent);

\node[native, right=3cm of parent]
(caelement){CommutativeAlgebraElement};

\node[diff, below=1.25cm of caelement](scalarfield)
{ScalarField\\ \scriptsize{\it parent:} ScalarFieldAlgebra};

\path[line] (scalarfield) -- (caelement);

\node[diff, below=of scalarfield](zeroscalarfield)
{ZeroScalarField\\ \scriptsize {\it parent:} ScalarFieldAlgebra};

\path[line] (zeroscalarfield) -- (scalarfield);


\node[native_legend, left=8cm of zeroscalarfield]
(native_legend){};
\node[empty, right=0.5em of native_legend]
{Native \Sage{} class};

\node[diff_legend, below=1.em of native_legend]
(diff_legend){};
\node[empty, right=0.5em of diff_legend]
{\SM{} class\\ \footnotesize (differential part)};

\end{tikzpicture}
\end{center}
\caption{\label{f:scalar_classes} Python classes for scalar fields
on a manifold.}
\end{figure}
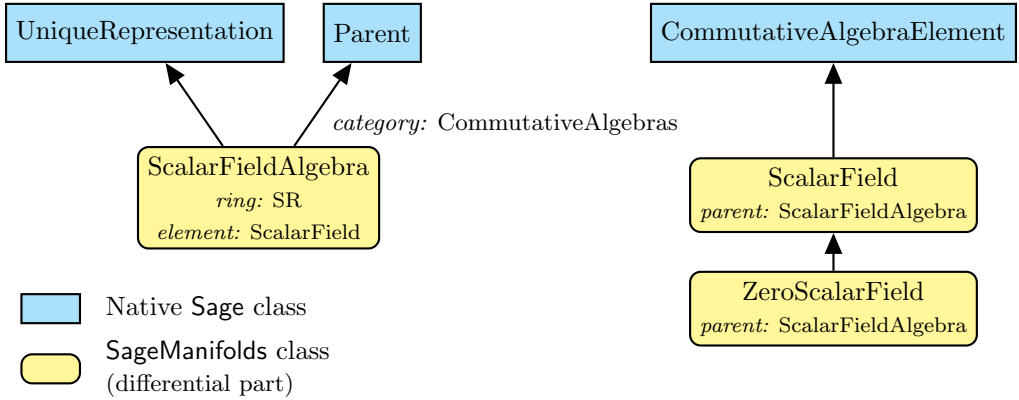

Given an open subset $U\subset\mathcal{M}$, the set $C^\infty(U)$
of scalar fields defined on $U$ has naturally the structure of a 
commutative algebra over $\mathbb{R}$: it is clearly a vector
space over $\mathbb{R}$ and it is endowed with a commutative ring structure
by pointwise multiplication:
\be
\forall f, g \in C^\infty(U),\quad \forall p\in U,\quad
(f.g)(p) := f(p) g(p) .
\ee
The algebra $C^\infty(U)$ is implemented in \SM{} via the parent
class \code{ScalarFieldAlgebra}, in the category
\code{CommutativeAlgebras}. The corresponding element class 
is of course \code{ScalarField} (cf. Fig.~\ref{f:scalar_classes}). 

\subsection{Modules and free modules} \label{s:modules}

Given an open subset $U\subset\mathcal{M}$, the set $\mathscr{X}(U)$
of all smooth vector fields defined on $U$ has naturally the structure of a 
\emph{module over the algebra} $C^\infty(U)$.
Let us recall that a \emph{module} is similar to a \emph{vector space}, except that it is based
on a \emph{ring} (here $C^\infty(U)$)
instead of a \emph{field} (usually $\mathbb{R}$ or
$\mathbb{C}$ in physical applications). Of course, every vector space is a module,
since every field is a ring.
There is an important difference though: every vector space has a basis (as a 
consequence of the axiom of choice),
while a module does not necessarily have any. 
When it possesses one, it is called a \emph{free module}. 
Moreover, if the module's base ring is commutative, it can be shown that all
bases have the same
cardinality, which is called the \emph{rank} of the module 
(for vector spaces, which are free modules, the word \emph{dimension}
is used instead of \emph{rank}). 

If $\mathscr{X}(U)$ is a free module (examples are provided in Sec.~\ref{s:vector_fields} below),
a basis of it is nothing but a \emph{vector frame}
$(\w{e}_a)_{1\leq a \leq n}$ on $U$
(often called a \emph{tetrad} in the context of 4-dimensional GR):
\be \label{e:v_expand}
    \forall \w{v}\in\mathscr{X}(U),\quad \w{v} = v^a \w{e}_a,\quad\mbox{with\ } v^a \in C^\infty(U) .
\ee
The rank of $\mathscr{X}(U)$ is thus $n$, i.e. the manifold's 
dimension\footnote{Note that the dimensionality of $\mathscr{X}(U)$ depends
of the adopted structure: as a vector space over $\mathbb{R}$, the
dimension of $\mathscr{X}(U)$ is infinite, while as a free module over
$C^\infty(U)$, $\mathscr{X}(U)$ has a finite rank. Note also that if $\mathscr{X}(U)$
is not free (i.e. no global vector frame exists on $U$), the notion of rank 
is meaningless.}.
At any point $p\in U$, Eq.~(\ref{e:v_expand}) gives birth to an identity in 
the tangent vector space $T_p \mathcal{M}$:
\be 
    \w{v}(p) = v^a(p)  \; \w{e}_a(p),\quad\mbox{with\ } v^a(p) \in \mathbb{R} , 
\ee
which means that 
the set $(\w{e}_a(p))_{1\leq a \leq n}$ is a basis of $T_p \mathcal{M}$.
Note that if $U$ is covered by a chart $(x^a)_{1\leq a \leq n}$,
then $(\partial/\partial x^a)_{1\leq a \leq n}$
is a vector frame on $U$, usually called \emph{coordinate frame}
or \emph{natural basis}. Note also that, being a
vector space over $\mathbb{R}$, the tangent space $T_p \mathcal{M}$
represents another kind of free module which occurs naturally in the
current context.

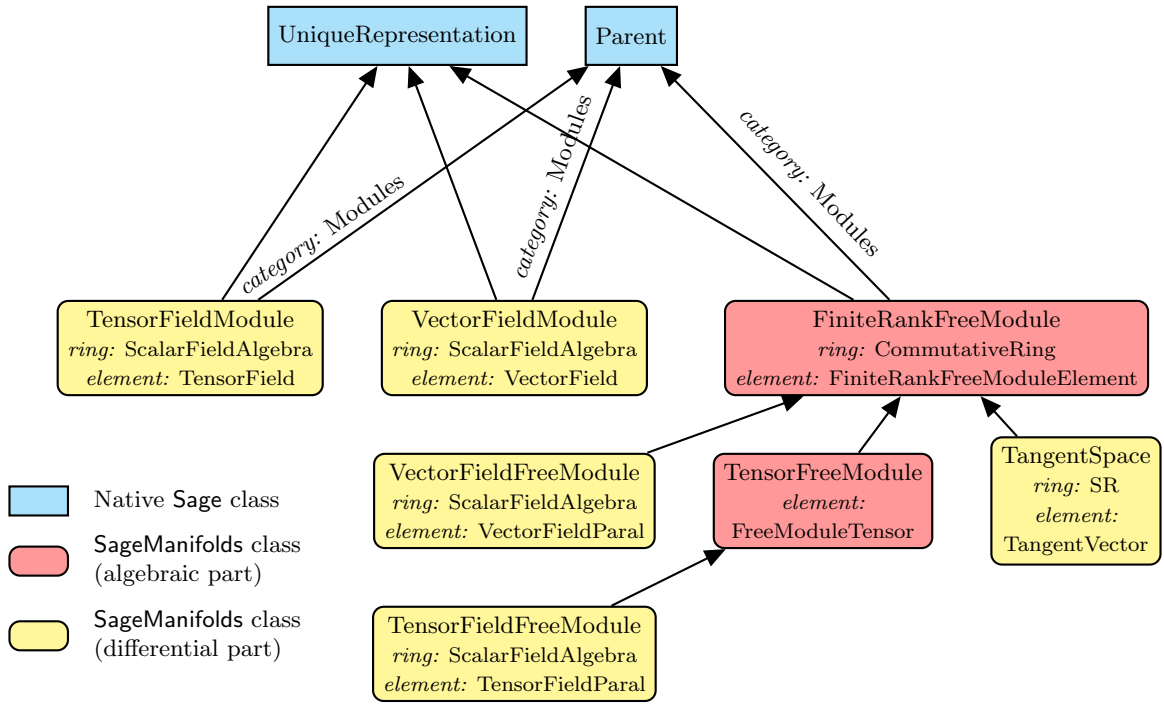
\begin{figure}
\begin{center}
\begin{tikzpicture}[font=\footnotesize, node distance=0.75cm, minimum
height=2em, auto]

\node[native]
(unique_represenation){UniqueRepresentation};

\node[native, right=of unique_represenation]
(parent){Parent};

\coordinate (Middle) at ($(unique_represenation)!0.5!(parent)$);

\node[diff, below=3.5cm of Middle]
(vectorfieldmodule)
{VectorFieldModule\\
{\scriptsize {\it ring:} ScalarFieldAlgebra}\\
{\scriptsize {\it element:} VectorField}};

\path[line] (vectorfieldmodule) -- (unique_represenation);
\path[line] (vectorfieldmodule) -- node [near start, yshift=5em,
xshift=2.2em, rotate=70]
{\footnotesize {\it category:} Modules} (parent);

\node[diff, left=of vectorfieldmodule]
(tensorfieldmodule)
{TensorFieldModule\\
{\scriptsize {\it ring:} ScalarFieldAlgebra}\\
{\scriptsize {\it element:} TensorField}};

\path[line] (tensorfieldmodule) -- (unique_represenation);
\path[line] (tensorfieldmodule) -- node [near start, xshift=3em,
yshift=1.5em, rotate=35]
{\footnotesize {\it category:} Modules} (parent);

\node[diff, below=of vectorfieldmodule](vectorfieldfreemodule)
{VectorFieldFreeModule\\
{\scriptsize {\it ring:} ScalarFieldAlgebra}\\
{\scriptsize {\it element:} VectorFieldParal}};

\node[diff, below=of vectorfieldfreemodule](tensorfieldfreemodule)
{TensorFieldFreeModule\\
{\scriptsize {\it ring:} ScalarFieldAlgebra}\\
{\scriptsize {\it element:} TensorFieldParal}};

\node[alg, right=1cm of vectorfieldmodule](finiterankfreemodule)
{FiniteRankFreeModule\\
{\scriptsize {\it ring:} CommutativeRing}\\
{\scriptsize {\it element:} FiniteRankFreeModuleElement}};

\node[alg, right=of vectorfieldfreemodule](tensorfreemodule)
{TensorFreeModule\\
{\scriptsize {\it element:}}\\
{\scriptsize FreeModuleTensor}};

\node[diff, right=of tensorfreemodule](tangentspace)
{TangentSpace\\
{\scriptsize {\it ring:} SR}\\
{\scriptsize {\it element:}}\\
{\scriptsize TangentVector}};

\path[line] (finiterankfreemodule) -- (unique_represenation);
\path[line] (finiterankfreemodule) -- node [near start,
xshift=2.5em, rotate=-47]
{\footnotesize {\it category:} Modules} (parent);

\path[line] (tangentspace) -- (finiterankfreemodule);
\path[line] (tensorfreemodule) -- (finiterankfreemodule);

\path[line] (vectorfieldfreemodule) -- (finiterankfreemodule);
\path[line] (tensorfieldfreemodule) -- (tensorfreemodule);

\node[native_legend, left=4cm of vectorfieldfreemodule]
(native_legend){};
\node[empty, right=0.5em of native_legend]
{Native \Sage{} class};

\node[alg_legend, below=1.em of native_legend]
(alg_legend){};
\node[empty, right=0.5em of alg_legend]
{\SM{} class\\ \footnotesize (algebraic part)};

\node[diff_legend, below=1.5em of alg_legend]
(diff_legend){};
\node[empty, right=0.5em of diff_legend]
{\SM{} class\\ \footnotesize (differential part)};

\end{tikzpicture}
\end{center}
\caption{\label{f:module_classes} Python classes for modules. For each of them,
the class of the base ring is indicated, as well as
the class for the elements.}
\end{figure}

It turns out that so far only free modules \emph{with a distinguished basis} were 
implemented in \Sage{}. This means that, given a free module $M$ of rank $n$, 
all calculations refer to a single basis of $M$. This amounts to identifying
$M$ with $R^n$, where $R$ is the ring over which $M$ is based. 
This is unfortunately not sufficient for dealing with smooth manifolds in a
coordinate-independent way. 
For instance, there is no canonical 
isomorphism between $T_p\mathcal{M}$ and $\mathbb{R}^n$ when no coordinate
system is privileged in the neighborhood of $p$.
Therefore we have started a pure algebraic part of \SM{} to implement
generic free modules, with an arbitrary number of bases, 
none of them being distinguished. This resulted in (i) the parent class 
\code{FiniteRankFreeModule}, within \Sage{}'s category \code{Modules}, and (ii)
the element class 
\code{FiniteRankFreeModuleElement}. Then both classes
\code{VectorFieldFreeModule} (for $\mathscr{X}(U)$, when it is 
a free module) and \code{TangentSpace} (for $T_p\mathcal{M}$)
inherit from \code{FiniteRankFreeModule} (see Fig.~\ref{f:module_classes}). 

\subsection{Implementation of vector fields} \label{s:vector_fields}

Ultimately, in \SM{}, vector fields are described by their 
components with respect to various vector frames, according to 
Eq.~(\ref{e:v_expand}), but without any vector frame being privileged, 
leaving the freedom to select one to the user, as well as to change 
coordinates. A key point is that 
not every manifold admits a global vector frame. 
A manifold $\mathcal{M}$, or more generally an open subset $U\subset\mathcal{M}$,
that admits a global vector frame is called 
\emph{parallelizable}. Equivalently,
$\mathcal{M}$ is parallelizable if, and only if, $\mathscr{X}(\mathcal{M})$
is a free module. In terms of tangent bundles, 
parallelizable manifolds are those for which the tangent bundle is trivial:
$T\mathcal{M} \simeq \mathcal{M}\times \mathbb{R}^n$.
Examples of parallelizable manifolds are \cite{Lee13}
\begin{itemize}
\item the Cartesian space $\mathbb{R}^n$ for $n=1,2,\ldots$, 
\item the circle $\mathbb{S}^1$, 
\item the torus $\mathbb{T}^2 = \mathbb{S}^1\times \mathbb{S}^1$, 
\item the sphere $\mathbb{S}^3 \simeq \mathrm{SU}(2)$, as any Lie group, 
\item the sphere $\mathbb{S}^7$, 
\item any orientable 3-manifold (Steenrod theorem \cite{Steen51}).
\end{itemize}
On the other hand, examples of non-parallelizable manifolds are
\begin{itemize}
\item the sphere $\mathbb{S}^2$ (as a consequence of the hairy ball theorem),
as well as any sphere $\mathbb{S}^n$ with $n\not\in\{1,3,7\}$, 
\item the real projective plane $\mathbb{RP}^2$.
\end{itemize}
Actually, ``most'' manifolds are non-parallelizable. 
As noticed above, if a manifold is covered by a single chart, it is 
parallelizable (the prototype being $\mathbb{R}^n$). But the reverse is not 
true: $\mathbb{S}^1$ and $\mathbb{T}^2$ are parallelizable and require 
at least two charts to cover them. 

\begin{figure}
\begin{center}
\begin{tikzpicture}[font=\small, node distance=0.5cm, minimum
height=2em, auto]

\node[diff](tensorfield)
{TensorField\\ \scriptsize {\it parent:}\\ \scriptsize
TensorFieldModule};

\node[diff, below=of tensorfield](vectorfield)
{VectorField\\ \scriptsize {\it parent:}\\ \scriptsize
VectorFieldModule};

\path[line] (vectorfield) -- (tensorfield);

\node[diff, right=of vectorfield](tensorfieldparal)
{TensorFieldParal\\ \scriptsize {\it parent:}\\ \scriptsize
TensorFieldFreeModule};

\node[diff, below=of tensorfieldparal](Vectorfieldparal)
{VectorFieldParal\\ \scriptsize {\it parent:}\\ \scriptsize
VectorFieldFreeModule};

\path[line] (Vectorfieldparal) -- (tensorfieldparal);
\path[line] (Vectorfieldparal) -- (vectorfield);
\path[line] (tensorfieldparal) -- (tensorfield);

\node[alg, right=6cm of tensorfield](freemoduletensor)
{FreeModuleTensor\\ \scriptsize {\it parent:}\\ \scriptsize
TensorFreeModule};

\node[alg, below=of freemoduletensor](finiterankfreemoduleelement)
{FiniteRankFreeModuleElement\\ \scriptsize {\it parent:}\\ \scriptsize
FiniteRankFreeModule};

\node[diff, below=of finiterankfreemoduleelement](tangentvector)
{TangentVector\\ \scriptsize {\it parent:}\\ \scriptsize
TangentSpace};

\node[native, above=4cm of tensorfieldparal]
(element){Element};

\node[native, below=of element]
(module_element){ModuleElement\\ {\scriptsize {\it parent:} Module}};

\path[line] (module_element) -- (element);

\path[line] (tensorfield) -- (module_element);

\path[line] (finiterankfreemoduleelement) -- (freemoduletensor);
\path[line] (Vectorfieldparal) -- (finiterankfreemoduleelement);

\path[line] (tangentvector) -- (finiterankfreemoduleelement);

\path[line] (tensorfieldparal) -- (freemoduletensor);

\path[line] (freemoduletensor) -- (module_element);

\node[native_legend, left=4cm of Vectorfieldparal]
(native_legend){};
\node[empty, right=0.5em of native_legend]
{Native \Sage{} class};

\node[alg_legend, below=1.em of native_legend]
(alg_legend){};
\node[empty, right=0.5em of alg_legend]
{\SM{} class\\ \footnotesize (algebraic part)};

\node[diff_legend, below=1.5em of alg_legend]
(diff_legend){};
\node[empty, right=0.5em of diff_legend]
{\SM{} class\\ \footnotesize (differential part)};

\end{tikzpicture}
\end{center}
\caption{\label{f:tensor_classes} Python classes implementing tensors and tensor fields.}
\end{figure}
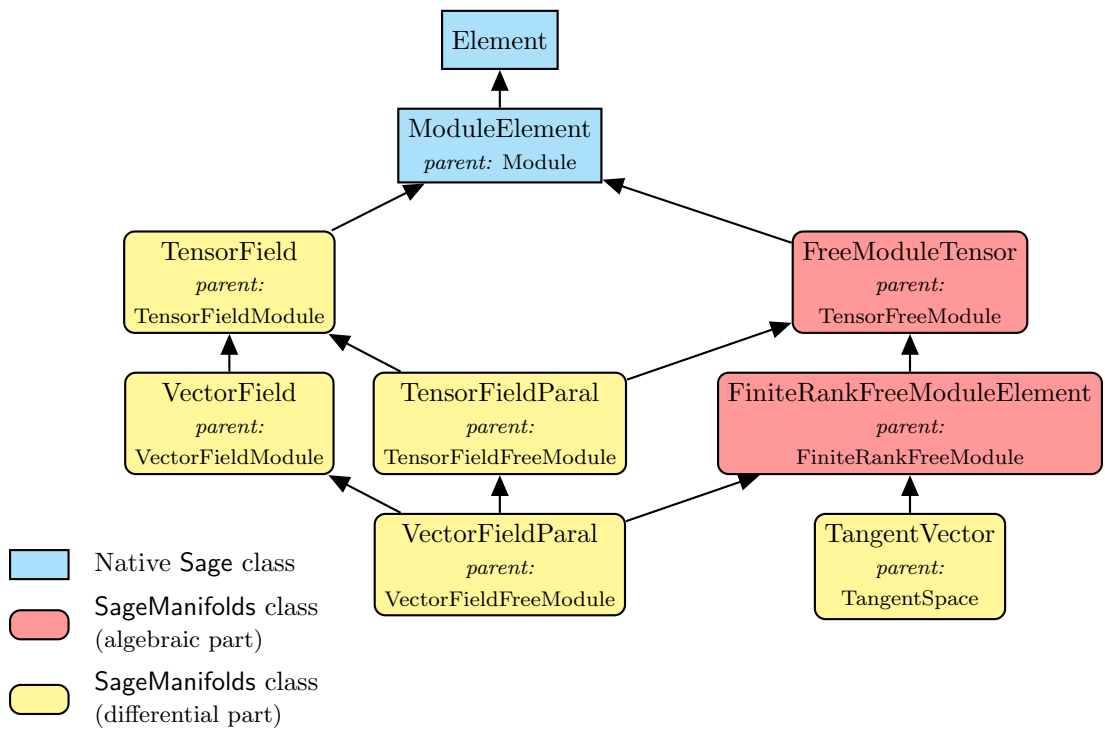

If the manifold $\mathcal{M}$ is not parallelizable,
we assume that it can be covered by a finite number $N$
of parallelizable open subsets $U_i$ ($1\leq i \leq N$).
In particular, this holds if $\mathcal{M}$ is compact, for any compact
manifold admits a finite atlas. 
We then consider the restrictions of vector fields to the $U_i$'s.
For each $i$, $\mathscr{X}(U_i)$ is a free module of rank $n=\mathrm{dim}\, \mathcal{M}$ and is implemented in \SM{} as an instance of 
\code{VectorFieldFreeModule} (cf. Sec.~\ref{s:modules} and Figs.~\ref{f:module_classes}
and \ref{f:tensor_classes}). 
Each vector field $\w{v}\in  \mathscr{X}(U_i)$ has different sets
of components $(v^a)_{1\leq a\leq n}$ in different vector frames 
$(\w{e}_a)_{1\leq a \leq n}$ introduced on $U_i$ [cf. Eq.~(\ref{e:v_expand})]. They are stored
as a Python dictionary whose keys are the vector frames:
\be
\w{v}.\mbox{\texttt{\_components}} = \left\{ (\w{e}): (v^a),
\ (\w{\hat e}): ({\hat v}^a), \ldots \right\}. 
\ee

\subsection{Implementation of tensor fields}

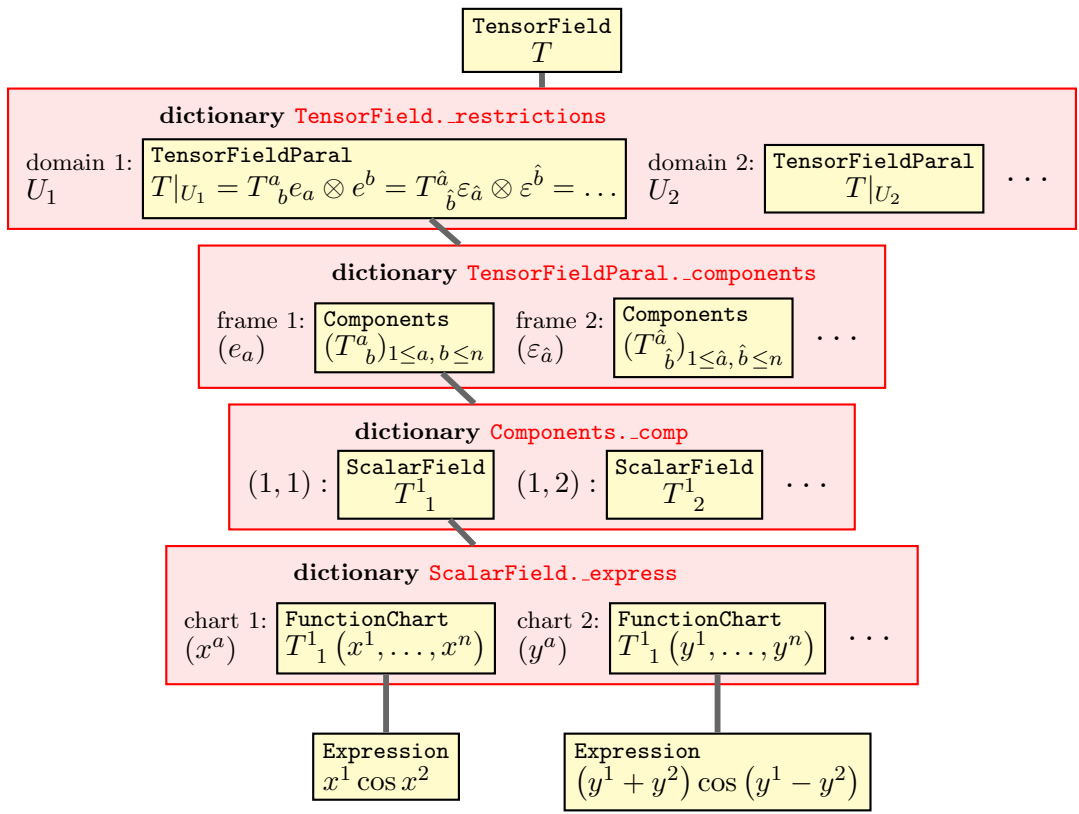
\begin{figure}
\begin{center}
\begin{tikzpicture}[font=\footnotesize, remember picture,
node distance=0.5em, minimum height=0.5em, auto]

\node[tens, align=center](tensorfield)
{\code{TensorField}\\ \normalsize $T$};

\node[dict, below=of tensorfield](restrictions){
\begin{tikzpicture}
  \node[empty, align=left](description)
  {\bf dictionary \textcolor{red}{\code{TensorField.\_restrictions}}};
  \node[empty, below left=of description](domain1)
  {domain 1:\\ {\normalsize $U_1$}};
  \node[tens, right=0em of domain1](tensorfield1)
  {\code{TensorFieldParal}\\
  {\normalsize $T|_{U_1}=T^a_{\ \, b} e_a\otimes
e^b=T^{\hat{a}}_{\ \, \hat{b}}\varepsilon_{\hat{a}}\otimes
\varepsilon^{\hat{b}}=\dots$}};
  \node[empty, right=of tensorfield1](domain2)
  {domain 2:\\ {\normalsize $U_2$}};
  \node[tens, align=center, right=0em of domain2](tensorfield2)
  {\code{TensorFieldParal}\\ {\normalsize $T|_{U_2}$}};
  \node[empty, right=of tensorfield2](more)
  {\large $\dots$};
\end{tikzpicture}
};

\node[dict, below=of restrictions](components){
\begin{tikzpicture}
  \node[empty, align=left](description)
  {\bf dictionary
\textcolor{red}{\code{TensorFieldParal.\_components}}};
  \node[empty, below left=of description](frame1)
  {frame 1:\\ {\normalsize $(e_a)$}};
  \node[tens, right=0em of frame1](components1)
  {\code{Components}\\
  {\normalsize $(T^a_{\ \, b})_{1\le a,\,b\,\le n}$}};
  \node[empty, right=of components1](frame2)
  {frame 2:\\ {\normalsize $(\varepsilon_{\hat{a}})$}};
  \node[tens, right=0em of frame2](components2)
  {\code{Components}\\
  {\normalsize $(T^{\hat{a}}_{\ \, \hat{b}})_{1\le
\hat{a},\,\hat{b}\,\le n}$}};
  \node[empty, right=of components2](more)
  {\large $\dots$};
\end{tikzpicture}
};

\node[dict, below=of components](comp){
\begin{tikzpicture}
  \node[empty, align=left](description)
  {\bf dictionary \textcolor{red}{\code{Components.\_comp}}};
  \node[empty, below left=of description](comp1)
  {\normalsize $(1,1):$};
  \node[tens, align=center, right=0em of comp1](scalarfield1)
  {\code{ScalarField}\\
  {\normalsize $T^1_{\ \, 1}$}};
  \node[empty, right=of scalarfield1](comp2)
  {\normalsize $(1,2):$};
  \node[tens, align=center, right=0em of comp2](scalarfield2)
  {\code{ScalarField}\\
  {\normalsize $T^1_{\ \, 2}$}};
  \node[empty, right=of scalarfield2](more)
  {\large $\dots$};
\end{tikzpicture}
};

\node[dict, below=of comp](express){
\begin{tikzpicture}
  \node[empty, align=left](description)
  {\bf dictionary \textcolor{red}{\code{ScalarField.\_express}}};
  \node[empty, below left=of description](chart1)
  {chart 1:\\
  {\normalsize $\left(x^a\right)$}};
  \node[tens, right=0em of chart1](functionchart1)
  {\code{FunctionChart}\\
  {\normalsize $T^1_{\ \, 1}\left(x^1,\dots,x^n\right)$}};
  \node[empty, right=of functionchart1](chart2)
  {chart 2:\\
  {\normalsize $\left(y^a\right)$}};
  \node[tens, right=0em of chart2](functionchart2)
  {\code{FunctionChart}\\
  {\normalsize $T^1_{\ \, 1}\left(y^1,\dots,y^n\right)$}};
  \node[empty, right=of functionchart2](more)
  {\large $\dots$};
\end{tikzpicture}
};

\node[tens, below=2em of functionchart1](express1)
{\code{Expression}\\
{\normalsize $x^1\cos x^2$}};
\node[tens, below=2em of functionchart2](express2)
{\code{Expression}\\
{\normalsize $\left(y^1 + y^2\right)\cos\left(y^1 - y^2\right)$}};

\draw[thick, line width=0.2em, black!60,-] (restrictions) --
(tensorfield);
\draw[thick, line width=0.2em, black!60,-] (components) --
(tensorfield1);
\draw[thick, line width=0.2em, black!60,-] (comp) -- (components1);
\draw[thick, line width=0.2em, black!60,-] (express) -- (scalarfield1);
\draw[thick, line width=0.2em, black!60,-] (express1) --
(functionchart1);
\draw[thick, line width=0.2em, black!60,-] (express2) --
(functionchart2);

\end{tikzpicture}
\end{center}
\caption{\label{f:tensorfield_structure} Storage of tensor fields in \SM{}.
Each red box represents a Python dictionary; the dictionary values are
depicted by yellow boxes, the keys being indicated at the left of each
box.}
\end{figure}

The implementation of tensor fields in \SM{} follows the strategy 
adopted for vector fields. Consider for instance a tensor field $\w{T}$ 
of type (1,1) on the manifold $\mathcal{M}$. 
It can be represented by 
components $T^a_{\ \, b}$ only on a parallelizable open subset $U\subset
\mathcal{M}$, since the decomposition 
\be \label{e:T_expand}
    \left. \w{T} \right| _{U} = T^a_{\ \, b} \, \w{e}_a \otimes \w{e}^b ,
\ee
which defines $T^a_{\ \, b}$, is meaningful only when a vector frame 
$(\w{e}_a)$ exists\footnote{Using standard notation, in Eq.~(\ref{e:T_expand}),
$(\w{e}^b)$ stands for the coframe dual to $(\w{e}_a)$}.
Therefore, one first decomposes the tensor field $\w{T}$  into 
its restrictions $\left. \w{T} \right| _{U_i}$ on parallelizable open subsets
of $\mathcal{M}$, $U_i$ ($1\leq i\leq N$) and then considers 
the components on various vector frames on each subset $U_i$.
For each vector frame $(\w{e}_a)$, the set of components $(T^a_{\ \, b})$
is stored in a devoted class (named \code{Components}), which takes into account
all the tensor monoterm symmetries: only non-redundant
components are stored, the other ones being deduced by (anti)symmetry. 
This is illustrated in Fig.~\ref{f:tensorfield_structure}, which depicts the
internal storage of tensor fields in \SM{}. Note that each component
$T^a_{\ \, b}$ is a scalar field on $U_i$, according to the formula
\be
    T^a_{\ \, b} = \w{T}(\w{e}^a, \w{e}_b) . 
\ee
Accordingly, the penultimate level of Fig.~\ref{f:tensorfield_structure} 
corresponds to the scalar field storage, as described by (\ref{e:f_express}).
The last level is constituted by \Sage{}'s symbolic expressions
(class \code{Expression}). 

\section{Current status of SageManifolds}

\subsection{Functionalities} \label{s:functionalities}
At present (version 0.6), the functionalities included in \SM{} are as
follows:  
\begin{itemize}
\item maps between manifolds and  pullback operator,
\item submanifolds and pushforward operator,
\item standard tensor calculus (tensor product, 
contraction, symmetrization, etc.), even on non-parallelizable manifolds,
\item arbitrary monoterm tensor symmetries, 
\item exterior calculus (wedge product and exterior derivative, Hodge duality),
\item Lie derivatives along a vector field,
\item affine connections (curvature, torsion),
\item pseudo-Riemannian metrics (Levi-Civita connection, Weyl tensor),
\item graphical display of charts.
\end{itemize}

\subsection{Parallelization}


To improve the reactivity of \SM{} and take advantage of
multicore processors, some tensorial operations are performed by
parallel processes. The parallelization is implemented by means of
the Python library \code{multiprocessing}, via the built-in
\Sage{} decorator \code{@parallel}. Using it permits to define a function
that is run on different sub-processes. If $n$ processes are used, given a function
and a list of arguments for it, any process will call the function
with an element of the list, one at time, spanning all the list.

Currently\footnote{in the development version of \SM{}; this
will become available in version 0.7 of the stable release.},
the parallelized operations are tensor algebra, tensor
contractions, computation of the connection coefficient
and computation of Riemann tensor.

The parallelization of an operation is achieved by first creating a
function which computes the required operation on a subset of the
components of a tensor; second, by creating a list of $2n$ (twice the
number of used processes) arguments for this function. Then applying this function
to the input list, the calculation is performed in parallel. 
At the end of the computation a fourth phase is needed to retrieve the
results.
The choice to divide the work in $2n$ is a compromise between the 
load balancing and the cost of creating multiple processes.
The number of processors to be used in the parallelization can be controlled by the
user.


\section{SageManifolds at work: Kerr spacetime and Simon-Mars tensor}

\begin{figure}
\begin{center}
\includegraphics[width=0.5\textwidth]{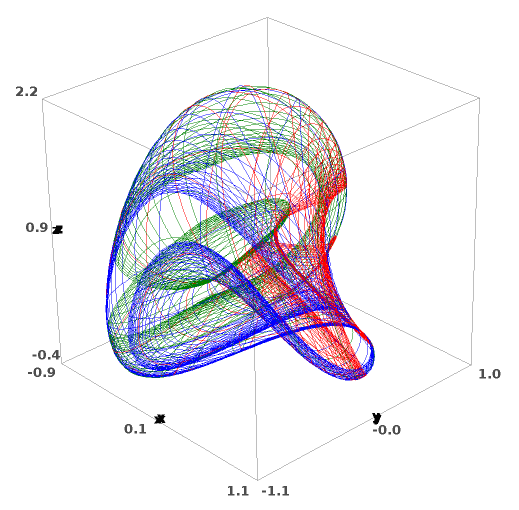}
\end{center}
\caption{\label{f:boy} Boy surface depicted via the grids of
3 coordinate charts covering
$\mathbb{RP}^2$ (see the text for the color code).}
\end{figure}

We give hereafter a short illustration of \SM{} focused on tensor 
calculus in 4-dimensional GR.
Another example, to be found at \cite{SM_examples}, is
based on the manifold $\mathbb{S}^2$ and 
focuses more on the use of multiple charts and on the treatment of non-parallelizable 
manifolds. Yet another example illustrates some graphical capabilities of
\SM{}: Figure~\ref{f:boy} shows the famous
immersion of the real projective plane $\mathbb{RP}^2$ into
the Euclidean space $\mathbb{R}^3$ known as the \emph{Boy surface}.
This figure has been obtained by means of the method \code{plot()} applied
to three coordinate charts covering $\mathbb{RP}^2$,
the definition of which is related to the interpretation of
$\mathbb{RP}^2$ as the set of straight lines $\Delta$
through the origin of $\mathbb{R}^3$:
(i) in red, the chart $X_1$ covering the open subset of $\mathbb{RP}^2$ defined
by all lines $\Delta$ that are not parallel to the plane $z=0$,
the coordinates of $X_1$ being the coordinates $(x,y)$ of the intersection of
the considered line $\Delta$ with the plane $z=1$;
(ii) in green, the chart $X_2$ covering the open subset defined
by all lines $\Delta$ that are not parallel to the plane $x=0$,
the coordinates of $X_2$ being the coordinates $(y,z)$ of intersection  with the plane $x=1$;
(iii) in blue, the chart $X_3$ covering the open subset defined
by all lines $\Delta$ that are not parallel to the plane $y=0$,
the coordinates of $X_2$ being the coordinates $(z,x)$ of intersection with the plane $y=1$.
Figure~\ref{f:boy} actually shows the coordinate grids of these three
charts through the Ap\'ery map \cite{Apery86}, which realizes an immersion
of $\mathbb{RP}^2$ into $\mathbb{R}^3$. This example, as many others,
can be found at \cite{SM_examples}.

Let us consider a 4-dimensional spacetime, i.e. a smooth 
4-manifold $\mathcal{M}$ endowed with a Lorentzian metric $\w{g}$. 
We assume that $(\mathcal{M},\w{g})$ is stationary and denote by $\w{\xi}$ 
the corresponding Killing vector field. 
The \emph{Simon-Mars tensor w.r.t. $\w{\xi}$} is then
the type-(0,3) tensor field $\w{S}$ defined by \cite{Mars99}
\be \label{e:def_Simon-Mars}
S_{\alpha\beta\gamma} := 4 \mathcal{C}_{\mu\alpha\nu[\beta} \, \xi^\mu \xi^\nu \, \sigma_{\gamma]}
 + \gamma_{\alpha[\beta} \, \mathcal{C}_{\gamma]\rho\mu\nu} \, \xi^\rho \, \mathcal{F}^{\mu\nu} ,
\ee
where
\begin{itemize}
\item $\gamma_{\alpha\beta} := \lambda \, g_{\alpha\beta} + \xi_\alpha \xi_\beta$, 
with $\lambda := - \xi_\mu \xi^\mu$;
\item $\mathcal{C}_{\alpha\beta\mu\nu} := C_{\alpha\beta\mu\nu}
    + \frac{i}{2} \epsilon^{\rho\sigma}_{\ \ \, \mu\nu}\,  C_{\alpha\beta\rho\sigma} $,
with $C^\alpha_{\ \, \beta\mu\nu}$ being the Weyl curvature tensor and 
$\epsilon_{\alpha\beta\mu\nu}$ the Levi-Civita volume 4-form;
\item $\mathcal{F}_{\alpha\beta} := F_{\alpha\beta} + i\,  {}^*\!F_{\alpha\beta}$, 
with $F_{\alpha\beta} := \nabla_\alpha\xi_\beta$ (Killing 2-form) and 
${}^*\!F_{\alpha\beta} := \frac{1}{2} \epsilon^{\mu\nu}_{\ \ \, \alpha\beta} F_{\mu\nu}$;
\item $\sigma_\alpha := 2 \mathcal{F}_{\mu\alpha} \xi^\mu$ (Ernst 1-form).
\end{itemize}
The Simon-Mars tensor provides a nice characterization of Kerr spacetime,
according the following theorem proved by Mars \cite{Mars99}:
if $\w{g}$ satisfies the vacuum Einstein equation and $(\mathcal{M},\w{g})$
contains a stationary asymptotically flat end $\mathcal{M}^\infty$ such
that $\w{\xi}$ tends to a time translation at infinity in $\mathcal{M}^\infty$
and the Komar mass of $\w{\xi}$ in $\mathcal{M}^\infty$ is non-zero, then 
$\w{S} = 0$ if, and only if, $(\mathcal{M},\w{g})$ is locally isometric 
to a Kerr spacetime.

In what follows, we use \SM{} to compute the Simon-Mars tensor 
according to formula (\ref{e:def_Simon-Mars}) for the Kerr metric and check 
that we get zero (the ``if'' part of the above theorem). 
The corresponding worksheet can be downloaded from \\
\url{http://sagemanifolds.obspm.fr/examples/html/SM_Simon-Mars_Kerr.html}.\\
For the sake of clarity, let us recall that, 
as an object-oriented language, Python (and hence \Sage{}) makes use of 
the following postfix notation:
\begin{center}
\textcolor{blue}{\texttt{result}} = \textcolor{red}{\texttt{object}}\texttt{.}\texttt{function(}\textcolor{green}{\texttt{arguments}}\texttt{)}
\end{center}
In a functional language, this would correspond to 
\textcolor{blue}{\texttt{result}} = \texttt{function(}\textcolor{red}{\texttt{object}}\texttt{,}\textcolor{green}{\texttt{arguments}}\texttt{)}.
For instance, the Riemann tensor of a metric $\w{g}$ is obtained as 
\verb+riem = g.riemann()+
(in this case, there is no extra argument, hence the empty parentheses). 
With this in mind, let us proceed with the computation by means of 
\SM{}. In the text below, the blue color denotes the outputs as they appear in
the \Sage{} notebook (note that all outputs are automatically
\LaTeX{}-formatted by \Sage).

The first step is to declare the Kerr spacetime (or more precisely the part of the Kerr spacetime covered by Boyer-Lindquist coordinates) as a 4-dimensional manifold:
\begin{verbatim}
M = Manifold(4, 'M', latex_name=r'\mathcal{M}')
print M
\end{verbatim}
\soutput{4-dimensional manifold 'M'}
The standard Boyer-Lindquist coordinates $(t,r,\theta,\phi)$
are introduced by declaring a chart $X$ on $\mathcal{M}$, via the
method \code{chart()}, the argument of which is a string expressing the
coordinates names, their ranges (the default is $(-\infty,+\infty)$)
and their \LaTeX{} symbols:
\begin{verbatim}
X.<t,r,th,ph> = M.chart('t r:(0,+oo) th:(0,pi):\\theta ph:(0,2*pi):\\phi') 
print X ; X
\end{verbatim}
\soutput{chart (M, (t, r, th, ph))\\
$(\mathcal{M},(t, r, \theta, \phi))$}
We define next the Kerr metric $\w{g}$ by setting its components in the 
coordinate frame associated with Boyer-Lindquist coordinates.
Since the latter is the current manifold's default frame (being the only one defined
at this stage), we do not need to specify it when referring to the components
by their indices:
\begin{verbatim}
g = M.lorentz_metric('g')
m = var('m') ; a = var('a')
rho2 = r^2 + (a*cos(th))^2
Delta = r^2 -2*m*r + a^2
g[0,0] = -(1-2*m*r/rho2)
g[0,3] = -2*a*m*r*sin(th)^2/rho2
g[1,1], g[2,2] = rho2/Delta, rho2
g[3,3] = (r^2+a^2+2*m*r*(a*sin(th))^2/rho2)*sin(th)^2
g.view()
\end{verbatim}
\soutput{$\displaystyle g = \left( -\frac{a^{2} \cos\left(\theta\right)^{2} - 2 \, m r +
r^{2}}{a^{2} \cos\left(\theta\right)^{2} + r^{2}} \right) \mathrm{d}
t\otimes \mathrm{d} t + \left( -\frac{2 \, a m r
\sin\left(\theta\right)^{2}}{a^{2} \cos\left(\theta\right)^{2} + r^{2}}
\right) \mathrm{d} t\otimes \mathrm{d} \phi \ + $\\
$\displaystyle \left( \frac{a^{2}
\cos\left(\theta\right)^{2} + r^{2}}{a^{2} - 2 \, m r + r^{2}} \right)
\mathrm{d} r\otimes \mathrm{d} r + \left( a^{2}
\cos\left(\theta\right)^{2} + r^{2} \right) \mathrm{d} \theta\otimes
\mathrm{d} \theta + \left( -\frac{2 \, a m r
\sin\left(\theta\right)^{2}}{a^{2} \cos\left(\theta\right)^{2} + r^{2}}
\right) \mathrm{d} \phi\otimes \mathrm{d} t + \left( \frac{2 \, a^{2} m
r \sin\left(\theta\right)^{4} + {\left(a^{2} r^{2} + r^{4} +
{\left(a^{4} + a^{2} r^{2}\right)} \cos\left(\theta\right)^{2}\right)}
\sin\left(\theta\right)^{2}}{a^{2} \cos\left(\theta\right)^{2} + r^{2}}
\right) \mathrm{d} \phi\otimes \mathrm{d} \phi $}
The Levi-Civita connection $\w{\nabla}$ associated with $\w{g}$ is obtained
by the method \code{connection()}:
\begin{verbatim}
nab = g.connection() ; print nab
\end{verbatim}
\soutput{Levi-Civita connection 'nabla\_g' associated with the Lorentzian metric
'g' on the 4-dimensional manifold 'M'}
As a check, we verify that the covariant derivative of $\w{g}$ with respect to
$\w{\nabla}$ vanishes identically:
\begin{verbatim}
nab(g).view()
\end{verbatim}
\soutput{$ \nabla_{g} g = 0$}
As mentionned above, 
the default vector frame on the spacetime manifold is the coordinate basis associated with Boyer-Lindquist coordinates:
\begin{verbatim}
M.default_frame() is X.frame()
\end{verbatim}
\soutput{True}
\begin{verbatim}
X.frame()
\end{verbatim}
\soutput{$\left(\mathcal{M} ,\left(\frac{\partial}{\partial t
},\frac{\partial}{\partial r },\frac{\partial}{\partial \theta
},\frac{\partial}{\partial \phi }\right)\right)$}
Let us consider the first vector field of this frame:
\begin{verbatim}
xi = X.frame()[0] ; xi
\end{verbatim}
\soutput{$\frac{\partial}{\partial t}$}
\begin{verbatim}
print xi 
\end{verbatim}
\soutput{vector field 'd/dt' on the 4-dimensional manifold 'M'}
The 1-form associated to it by metric duality is
\begin{verbatim}
xi_form = xi.down(g)
xi_form.set_name('xi_form', r'\underline{\xi}')
print xi_form ; xi_form.view()
\end{verbatim}
\soutput{1-form 'xi\_form' on the 4-dimensional manifold 'M'\\[1ex]
$\underline{\xi} = \left( -\frac{a^{2} \cos\left(\theta\right)^{2} - 2 \,
m r + r^{2}}{a^{2} \cos\left(\theta\right)^{2} + r^{2}} \right)
\mathrm{d} t + \left( -\frac{2 \, a m r
\sin\left(\theta\right)^{2}}{a^{2} \cos\left(\theta\right)^{2} + r^{2}}
\right) \mathrm{d} \phi$}
Its covariant derivative is
\begin{verbatim}
nab_xi = nab(xi_form)
print nab_xi ; nab_xi.view()
\end{verbatim}
\soutput{tensor field 'nabla\_g xi\_form' of type (0,2) on the 4-dimensional
manifold 'M'\\[1ex]
$\nabla_{g} \underline{\xi} = \left( \frac{a^{2} m
\cos\left(\theta\right)^{2} - m r^{2}}{a^{4} \cos\left(\theta\right)^{4}
+ 2 \, a^{2} r^{2} \cos\left(\theta\right)^{2} + r^{4}} \right)
\mathrm{d} t\otimes \mathrm{d} r + \left( \frac{2 \, a^{2} m r
\cos\left(\theta\right) \sin\left(\theta\right)}{a^{4}
\cos\left(\theta\right)^{4} + 2 \, a^{2} r^{2}
\cos\left(\theta\right)^{2} + r^{4}} \right) \mathrm{d} t\otimes
\mathrm{d} \theta + $\\
$\left( -\frac{a^{2} m \cos\left(\theta\right)^{2} -
m r^{2}}{a^{4} \cos\left(\theta\right)^{4} + 2 \, a^{2} r^{2}
\cos\left(\theta\right)^{2} + r^{4}} \right) \mathrm{d} r\otimes
\mathrm{d} t + \left( \frac{{\left(a^{3} m \cos\left(\theta\right)^{2} -
a m r^{2}\right)} \sin\left(\theta\right)^{2}}{a^{4}
\cos\left(\theta\right)^{4} + 2 \, a^{2} r^{2}
\cos\left(\theta\right)^{2} + r^{4}} \right) \mathrm{d} r\otimes
\mathrm{d} \phi + $\\
$\left( -\frac{2 \, a^{2} m r \cos\left(\theta\right)
\sin\left(\theta\right)}{a^{4} \cos\left(\theta\right)^{4} + 2 \, a^{2}
r^{2} \cos\left(\theta\right)^{2} + r^{4}} \right) \mathrm{d}
\theta\otimes \mathrm{d} t + \left( \frac{2 \, {\left(a^{3} m r + a m
r^{3}\right)} \cos\left(\theta\right) \sin\left(\theta\right)}{a^{4}
\cos\left(\theta\right)^{4} + 2 \, a^{2} r^{2}
\cos\left(\theta\right)^{2} + r^{4}} \right) \mathrm{d} \theta\otimes
\mathrm{d} \phi + $ \\
$\left( -\frac{{\left(a^{3} m
\cos\left(\theta\right)^{2} - a m r^{2}\right)}
\sin\left(\theta\right)^{2}}{a^{4} \cos\left(\theta\right)^{4} + 2 \,
a^{2} r^{2} \cos\left(\theta\right)^{2} + r^{4}} \right) \mathrm{d}
\phi\otimes \mathrm{d} r + \left( -\frac{2 \, {\left(a^{3} m r + a m
r^{3}\right)} \cos\left(\theta\right) \sin\left(\theta\right)}{a^{4}
\cos\left(\theta\right)^{4} + 2 \, a^{2} r^{2}
\cos\left(\theta\right)^{2} + r^{4}} \right) \mathrm{d} \phi\otimes
\mathrm{d} \theta$}
Let us check that the Killing equation is satisfied:
\begin{verbatim}
nab_xi.symmetrize().view()
\end{verbatim}
\soutput{$0$}
Equivalently, we check that the Lie derivative of the metric along $\w{\xi}$
vanishes:
\begin{verbatim}
g.lie_der(xi).view()
\end{verbatim}
\soutput{$0$}
Thanks to Killing equation, $\w{\nabla} \underline{\w{\xi}}$ is antisymmetric. We may therefore define a 2-form by $\w{F}:= - \w{\nabla} \underline{\w{\xi}}$. Here we enforce the antisymmetry by calling the method \code{antisymmetrize()} on \code{nab\_xi}:
\begin{verbatim}
F = - nab_xi.antisymmetrize()
F.set_name('F')
print F ; F.view()
\end{verbatim}
\soutput{2-form 'F' on the 4-dimensional manifold 'M'\\[1ex]
$F = \left( -\frac{a^{2} m \cos\left(\theta\right)^{2} - m r^{2}}{a^{4}
\cos\left(\theta\right)^{4} + 2 \, a^{2} r^{2}
\cos\left(\theta\right)^{2} + r^{4}} \right) \mathrm{d} t\wedge
\mathrm{d} r + \left( -\frac{2 \, a^{2} m r \cos\left(\theta\right)
\sin\left(\theta\right)}{a^{4} \cos\left(\theta\right)^{4} + 2 \, a^{2}
r^{2} \cos\left(\theta\right)^{2} + r^{4}} \right) \mathrm{d} t\wedge
\mathrm{d} \theta + $\\
$\left( -\frac{{\left(a^{3} m
\cos\left(\theta\right)^{2} - a m r^{2}\right)}
\sin\left(\theta\right)^{2}}{a^{4} \cos\left(\theta\right)^{4} + 2 \,
a^{2} r^{2} \cos\left(\theta\right)^{2} + r^{4}} \right) \mathrm{d}
r\wedge \mathrm{d} \phi + \left( -\frac{2 \, {\left(a^{3} m r + a m
r^{3}\right)} \cos\left(\theta\right) \sin\left(\theta\right)}{a^{4}
\cos\left(\theta\right)^{4} + 2 \, a^{2} r^{2}
\cos\left(\theta\right)^{2} + r^{4}} \right) \mathrm{d} \theta\wedge
\mathrm{d} \phi$}
The squared norm of the Killing vector is
\begin{verbatim}
lamb = - g(xi,xi)
lamb.set_name('lambda', r'\lambda')
print lamb ; lamb.view()
\end{verbatim}
\soutput{scalar field 'lambda' on the 4-dimensional manifold 'M'\\[1ex]
$\begin{array}{llcl} \lambda:& \mathcal{M} & \longrightarrow
& \mathbb{R} \\ & \left(t, r, \theta, \phi\right) & \longmapsto
& \frac{a^{2} \cos\left(\theta\right)^{2} - 2 \, m r + r^{2}}{a^{2}
\cos\left(\theta\right)^{2} + r^{2}} \end{array}$\\[-1ex]}
Instead of invoking $\w{g}(\w{\xi},\w{\xi})$, we could have evaluated $\lambda$ 
by means of the 1-form $\w{\underline{\xi}}$ acting on the vector field $\w{\xi}$:
\begin{verbatim}
lamb == - xi_form(xi)
\end{verbatim}
\soutput{True}
or, using index notation as $\lambda = - \xi_a \xi^a$:
\begin{verbatim}
lamb == - ( xi_form['_a']*xi['^a'] )
\end{verbatim}
\soutput{True}
The Riemann curvature tensor associated with $\w{g}$ is
\begin{verbatim}
Riem = g.riemann()
print Riem
\end{verbatim}
\soutput{tensor field 'Riem(g)' of type (1,3) on the 4-dimensional manifold 'M'}
The component $R^0_{\ \, 123} = R^t_{\ \, r\theta\phi}$ is 
\begin{verbatim}
Riem[0,1,2,3]
\end{verbatim}
\soutput{$-\frac{{\left(a^{7} m - 2 \, a^{5} m^{2} r + a^{5} m r^{2}\right)}
\cos\left(\theta\right) \sin\left(\theta\right)^{5} + {\left(a^{7} m + 2
\, a^{5} m^{2} r + 6 \, a^{5} m r^{2} - 6 \, a^{3} m^{2} r^{3} + 5 \,
a^{3} m r^{4}\right)} \cos\left(\theta\right)
\sin\left(\theta\right)^{3}}{a^{2} r^{6} - 2 \, m r^{7} + r^{8} +
{\left(a^{8} - 2 \, a^{6} m r + a^{6} r^{2}\right)}
\cos\left(\theta\right)^{6} + 3 \, {\left(a^{6} r^{2} - 2 \, a^{4} m
r^{3} + a^{4} r^{4}\right)} \cos\left(\theta\right)^{4} + 3 \,
{\left(a^{4} r^{4} - 2 \, a^{2} m r^{5} + a^{2} r^{6}\right)}
\cos\left(\theta\right)^{2}}\\
+ \frac{2 \, {\left(a^{7} m - a^{5} m r^{2} - 5 \,
a^{3} m r^{4} - 3 \, a m r^{6}\right)} \cos\left(\theta\right)
\sin\left(\theta\right)}{a^{2} r^{6} - 2 \, m r^{7} + r^{8} +
{\left(a^{8} - 2 \, a^{6} m r + a^{6} r^{2}\right)}
\cos\left(\theta\right)^{6} + 3 \, {\left(a^{6} r^{2} - 2 \, a^{4} m
r^{3} + a^{4} r^{4}\right)} \cos\left(\theta\right)^{4} + 3 \,
{\left(a^{4} r^{4} - 2 \, a^{2} m r^{5} + a^{2} r^{6}\right)}
\cos\left(\theta\right)^{2}}$}
Let us check that the Kerr metric is 
a vacuum solution of Einstein equation, i.e. that the Ricci tensor
vanishes identically:
\begin{verbatim}
g.ricci().view()
\end{verbatim}
\soutput{$\mathrm{Ric}(g) = 0$}
The Weyl conformal curvature tensor is
\begin{verbatim}
C = g.weyl() ; print C
\end{verbatim}
\soutput{tensor field 'C(g)' of type (1,3) on the 4-dimensional manifold 'M'}
Let us exhibit the component $C^0_{\ \, 101}=C^t_{\ \, rtr}$:
\begin{verbatim}
C[0,1,0,1]
\end{verbatim}
\soutput{$\frac{3 \, a^{4} m r \cos\left(\theta\right)^{4} + 3 \, a^{2} m r^{3} +
2 \, m r^{5} - {\left(9 \, a^{4} m r + 7 \, a^{2} m r^{3}\right)}
\cos\left(\theta\right)^{2}}{a^{2} r^{6} - 2 \, m r^{7} + r^{8} +
{\left(a^{8} - 2 \, a^{6} m r + a^{6} r^{2}\right)}
\cos\left(\theta\right)^{6} + 3 \, {\left(a^{6} r^{2} - 2 \, a^{4} m
r^{3} + a^{4} r^{4}\right)} \cos\left(\theta\right)^{4} + 3 \,
{\left(a^{4} r^{4} - 2 \, a^{2} m r^{5} + a^{2} r^{6}\right)}
\cos\left(\theta\right)^{2}}$}
To form the Simon-Mars tensor, we need the fully covariant form (type-(0,4)
tensor) of the Weyl tensor (i.e. $C_{\alpha\beta\mu\nu} = g_{\alpha\sigma} C^\sigma_{\ \, \beta\mu\nu}$); we get it by lowering the first index with the metric:
\begin{verbatim}
Cd = C.down(g) ; print Cd
\end{verbatim}
\soutput{tensor field of type (0,4) on the 4-dimensional manifold 'M'}
The (monoterm) symmetries of this tensor are those inherited from the
Weyl tensor, i.e. the antisymmetry on the last two indices (position 2 and 3, the first index being at position 0):
\begin{verbatim}
Cd.symmetries()
\end{verbatim}
\soutput{no symmetry;  antisymmetry: (2, 3)}
Actually, \code{Cd} is also antisymmetric with respect to the first two indices (positions 0 and 1), as we can check:
\begin{verbatim}
Cd == Cd.antisymmetrize(0,1)
\end{verbatim}
\soutput{True}
To take this symmetry into account explicitly, we set
\begin{verbatim}
Cd = Cd.antisymmetrize(0,1)
Cd.symmetries()
\end{verbatim}
\soutput{no symmetry;  antisymmetries: [(0, 1), (2, 3)]}
The starting point in the evaluation of Simon-Mars tensor is the
self-dual complex 2-form associated with the Killing 2-form $\w{F}$, i.e. the object
$\w{\mathcal{F}} := \w{F} + i \, {}^* \w{F}$, where ${}^*\w{F}$ is the Hodge dual of $\w{F}$:
\begin{verbatim}
FF = F + I * F.hodge_star(g)
FF.set_name('FF', r'\mathcal{F}')
print FF ; FF.view()
\end{verbatim}
\soutput{2-form 'FF' on the 4-dimensional manifold 'M'\\
\small $\mathcal{F} = \left( -\frac{a^{2} m \cos\left(\theta\right)^{2} + 2 i \,
a m r \cos\left(\theta\right) - m r^{2}}{a^{4}
\cos\left(\theta\right)^{4} + 2 \, a^{2} r^{2}
\cos\left(\theta\right)^{2} + r^{4}} \right) \mathrm{d} t\wedge
\mathrm{d} r + \left( \frac{{\left(i \, a^{3} m
\cos\left(\theta\right)^{2} - 2 \, a^{2} m r \cos\left(\theta\right) - i
\, a m r^{2}\right)} \sin\left(\theta\right)}{a^{4}
\cos\left(\theta\right)^{4} + 2 \, a^{2} r^{2}
\cos\left(\theta\right)^{2} + r^{4}} \right) \mathrm{d} t\wedge
\mathrm{d} \theta\\ 
+ 
\left(\frac{-4 i \, a^{4} m^{2} r^{2}
\cos\left(\theta\right) \sin\left(\theta\right)^{4} + {\left(a^{3} m
r^{4} - 2 \, a m^{2} r^{5} + a m r^{6} - {\left(a^{7} m - 2 \, a^{5}
m^{2} r + a^{5} m r^{2}\right)} \cos\left(\theta\right)^{4} 
\right)} \sin\left(\theta\right)^{2}}
{a^{2} r^{6}
- 2 \, m r^{7} + r^{8} + {\left(a^{8} - 2 \, a^{6} m r + a^{6}
r^{2}\right)} \cos\left(\theta\right)^{6} + 3 \, {\left(a^{6} r^{2} - 2
\, a^{4} m r^{3} + a^{4} r^{4}\right)} \cos\left(\theta\right)^{4} + 3
\, {\left(a^{4} r^{4} - 2 \, a^{2} m r^{5} + a^{2} r^{6}\right)}
\cos\left(\theta\right)^{2}}\right.\\
-\left.\frac{
\left(
{\left(2 i
\, a^{6} m r + 2 i \, a^{4} m r^{3}\right)} \cos\left(\theta\right)^{3}
+{\left(-4 i \, a^{4} m^{2} r^{2} + 2 i \, a^{4} m r^{3} - 4 i \, a^{2}
m^{2} r^{4} + 2 i \, a^{2} m r^{5}\right)}
\cos\left(\theta\right)\right)\sin\left(\theta\right)^{2}}
{a^{2} r^{6}
- 2 \, m r^{7} + r^{8} + {\left(a^{8} - 2 \, a^{6} m r + a^{6}
r^{2}\right)} \cos\left(\theta\right)^{6} + 3 \, {\left(a^{6} r^{2} - 2
\, a^{4} m r^{3} + a^{4} r^{4}\right)} \cos\left(\theta\right)^{4} + 3
\, {\left(a^{4} r^{4} - 2 \, a^{2} m r^{5} + a^{2} r^{6}\right)}
\cos\left(\theta\right)^{2}}\right)\mathrm{d} r\wedge \mathrm{d} \phi\\
+ 
\left( -\frac{{\left(i \, a^{4} m + i \, a^{2} m r^{2}\right)}
\sin\left(\theta\right)^{3} + {\left(-i \, a^{4} m + i \, m r^{4} + 2 \,
{\left(a^{3} m r + a m r^{3}\right)} \cos\left(\theta\right)\right)}
\sin\left(\theta\right)}{a^{4} \cos\left(\theta\right)^{4} + 2 \, a^{2}
r^{2} \cos\left(\theta\right)^{2} + r^{4}} \right) \mathrm{d}
\theta\wedge \mathrm{d} \phi$}
Let us check that $\w{\mathcal{F}}$ is self-dual, i.e. that it obeys
${}^* \w{\mathcal{F}} = -i \w{\mathcal{F}}$:
\begin{verbatim}
FF.hodge_star(g) == - I * FF
\end{verbatim}
\soutput{True}
To evaluate the right self-dual of the Weyl tensor, we need the 
tensor $\epsilon^{\alpha\beta}_{\ \ \ \gamma\delta}$:
\begin{verbatim}
eps = g.volume_form(2)  # 2 = the first 2 indices are contravariant
print eps ; eps.symmetries()
\end{verbatim}
\soutput{tensor field of type (2,2) on the 4-dimensional manifold 'M'\\
no symmetry;  antisymmetries: [(0, 1), (2, 3)]}
The right self-dual Weyl tensor is then:
\begin{verbatim}
CC = Cd + I/2*( eps['^rs_..']*Cd['_..rs'] )
CC.set_name('CC', r'\mathcal{C}') ; 
print CC ; CC.symmetries()
\end{verbatim}
\soutput{tensor field 'CC' of type (0,4) on the 4-dimensional manifold 'M'\\
no symmetry;  antisymmetries: [(0, 1), (2, 3)]}
\begin{verbatim}
CC[0,1,2,3]
\end{verbatim}
\soutput{$\frac{{\left(a^{5} m \cos\left(\theta\right)^{5} + 3 i \, a^{4} m r
\cos\left(\theta\right)^{4} + 3 i \, a^{2} m r^{3} + 2 i \, m r^{5} -
{\left(3 \, a^{5} m + 5 \, a^{3} m r^{2}\right)}
\cos\left(\theta\right)^{3}\right)}
\sin\left(\theta\right)}
{a^{6} \cos\left(\theta\right)^{6} + 3 \, a^{4}
r^{2} \cos\left(\theta\right)^{4} + 3 \, a^{2} r^{4}
\cos\left(\theta\right)^{2} + r^{6}}\\
+\frac{{\left({\left(-9 i \, a^{4} m r - 7 i \, a^{2} m
r^{3}\right)} \cos\left(\theta\right)^{2} + 3 \, {\left(3 \, a^{3} m
r^{2} + 2 \, a m r^{4}\right)} \cos\left(\theta\right)\right)}
\sin\left(\theta\right)}
{a^{6} \cos\left(\theta\right)^{6} + 3 \, a^{4}
r^{2} \cos\left(\theta\right)^{4} + 3 \, a^{2} r^{4}
\cos\left(\theta\right)^{2} + r^{6}}$}
The Ernst 1-form $\sigma_\alpha = 2 \mathcal{F}_{\mu\alpha} \, \xi^\mu$ (0 = contraction on the first index of $\w{\mathcal{F}}$):
\begin{verbatim}
sigma = 2*FF.contract(0, xi)
\end{verbatim}
Instead of invoking the method \code{contract()}, 
we could have used the index notation to denote the contraction:
\begin{verbatim}
sigma == 2*( FF['_ma']*xi['^m'] )
\end{verbatim}
\soutput{True}
\begin{verbatim}
sigma.set_name('sigma', r'\sigma')
print sigma ; sigma.view()
\end{verbatim}
\soutput{1-form 'sigma' on the 4-dimensional manifold 'M'\\
$\sigma = \left( -\frac{2 \, a^{2} m \cos\left(\theta\right)^{2} + 4 i \,
a m r \cos\left(\theta\right) - 2 \, m r^{2}}{a^{4}
\cos\left(\theta\right)^{4} + 2 \, a^{2} r^{2}
\cos\left(\theta\right)^{2} + r^{4}} \right) \mathrm{d} r + \left(
\frac{{\left(2 i \, a^{3} m \cos\left(\theta\right)^{2} - 4 \, a^{2} m r
\cos\left(\theta\right) - 2 i \, a m r^{2}\right)}
\sin\left(\theta\right)}{a^{4} \cos\left(\theta\right)^{4} + 2 \, a^{2}
r^{2} \cos\left(\theta\right)^{2} + r^{4}} \right) \mathrm{d} \theta$}
The symmetric bilinear form
$\w{\gamma} = \lambda \, \w{g} + \underline{\w{\xi}}\otimes\underline{\w{\xi}}$:
\begin{verbatim}
gamma = lamb*g + xi_form * xi_form
gamma.set_name('gamma', r'\gamma')
print gamma ; gamma.view()
\end{verbatim}
\soutput{field of symmetric bilinear forms 'gamma' on the 4-dimensional manifold
'M'\\
$\gamma = \left( \frac{a^{2} \cos\left(\theta\right)^{2} - 2 \, m r +
r^{2}}{a^{2} - 2 \, m r + r^{2}} \right) \mathrm{d} r\otimes \mathrm{d}
r + \left( a^{2} \cos\left(\theta\right)^{2} - 2 \, m r + r^{2} \right)
\mathrm{d} \theta\otimes \mathrm{d} \theta +$\\
$\left( \frac{2 \, a^{2} m r
\sin\left(\theta\right)^{4} - {\left(2 \, a^{2} m r - a^{2} r^{2} + 2 \,
m r^{3} - r^{4} - {\left(a^{4} + a^{2} r^{2}\right)}
\cos\left(\theta\right)^{2}\right)} \sin\left(\theta\right)^{2}}{a^{2}
\cos\left(\theta\right)^{2} + r^{2}} \right) \mathrm{d} \phi\otimes
\mathrm{d} \phi$}
The first part of the Simon-Mars tensor is 
$S^{(1)}_{\alpha\beta\gamma} = 4 \mathcal{C}_{\mu\alpha\nu\beta} \, \xi^\mu \, \xi^\nu \, \sigma_\gamma$:
\begin{verbatim}
S1 = 4*( CC.contract(0,xi).contract(1,xi) ) * sigma
print S1
\end{verbatim}
\soutput{tensor field of type (0,3) on the 4-dimensional manifold 'M'}
The second part is 
$S^{(2)}_{\alpha\beta\gamma} = \gamma_{\alpha\beta} \, \mathcal{C}_{\rho\gamma\mu\nu} \, \xi^\rho \, \mathcal{F}^{\mu\nu}$, which we 
compute using the index notation to perform the contractions:
\begin{verbatim}
FFuu = FF.up(g)
xiCC = CC['_.r..']*xi['^r']
S2 = gamma * ( xiCC['_.mn']*FFuu['^mn'] )
print S2
\end{verbatim}
\soutput{tensor field of type (0,3) on the 4-dimensional manifold 'M'}
To get the Simon-Mars tensor, we need to antisymmetrize $\w{S}^{(1)}$ and
$\w{S}^{(2)}$ on their last two indices; we choose to use the standard
index notation to perform this operation (an alternative would have been
to call directly the method \code{antisymmetrize()}):
\begin{verbatim}
S1A = S1['_a[bc]']
S2A = S2['_a[bc]']
\end{verbatim}
The Simon-Mars tensor is then 
\begin{verbatim}
S = = S1A + S2A
S.set_name('S')
print S ; S.symmetries()
\end{verbatim}
\soutput{tensor field 'S' of type (0,3) on the 4-dimensional manifold 'M'\\
no symmetry;  antisymmetry: (1, 2)}
We check that it vanishes identically, as it should for Kerr spacetime:
\begin{verbatim}
S.view()
\end{verbatim}
\soutput{$S=0$}
\begin{verbatim}
\end{verbatim}
\begin{verbatim}
\end{verbatim}


\section{Conclusion and future prospects}

\SM{} is a work in progress. 
It encompasses currently $\sim 35,000$ lines of Python code (including comments and 
doctests). 
The last stable version (0.6), the functionalities of which are 
listed in Sec.~\ref{s:functionalities},
is freely downloadable from the project page
\cite{SM}. The development version (to become version 0.7 soon)
is also available from that page. Among future developments are 
\begin{itemize}
\item the extrinsic geometry of pseudo-Riemannian submanifolds,
\item the computation of geodesics (numerical integration via \soft{Sage/GSL} or 
\soft{Gyoto} \cite{Gyoto}),
\item evaluating integrals on submanifolds,
\item adding more graphical outputs,
\item adding more functionalities: symplectic forms, fibre bundles, 
spinors, variational calculus, etc.
\item the connection with numerical relativity.
\end{itemize}
The last point means using \SM{} for interactive exploration 
of numerically generated spacetimes. Looking at
the diagram in Fig.~\ref{f:tensorfield_structure}, one realizes that it 
suffices to replace only the lowest level, currently relying on
\Sage{}'s symbolic expressions, by computations on numerical data. 

Let us conclude by stating that, in the very spirit of free software, 
anybody interested in contributing to the project is very welcome!

\ack
This work has benefited from enlightening discussions with Volker Braun,
Vincent Delecroix, Simon King, S\'ebastien Labb\'e, Jos\'e M. Mart\'\i n-Garc\'\i a, 
Marc Mezzarobba, Thierry Monteil, Travis Scrimshaw,  Nicolas M.  Thi\'ery
and Anne Vaugon. 
We also thank St\'ephane M\'en\'e for his technical help and
Tolga Birkandan for fruitful comments about the manuscript.
EG acknowledges the warm hospitality of the Organizers of the
\emph{Encuentros Relativistas Espa\~noles 2014}, where this work was
presented.

\end{document}